\documentclass[paper ]{revtex4}

\usepackage{graphicx}
\usepackage{rotating}
\usepackage{amsmath}
\usepackage{url}
\usepackage{bm}

\newcommand{\bk}{{\bf k} }
\newcommand{\be}{\begin{equation}}
\newcommand{\ee}{\end{equation}}
\newcommand{\bea}{\begin{eqnarray}}
\newcommand{\eea}{\end{eqnarray}}
\newcommand{\ba}{\begin{array}}
\newcommand{\ea}{\end{array}}
\newcommand{\Ave}[1] { \langle #1 \rangle }
\newcommand{\aF}{\alpha^2 F(\Omega)}

\newcommand{\EF}{$E_F$}
\newcommand{\D}{$\Delta$}

\newcommand{\nk}{{n\bf k}}
\newcommand{\nkp}{{n'\bf k'}}

\newcommand{\br}{{\bf r}}
\newcommand{\bR}{{\bf R}}

\newcommand{\s}{$\sigma$}
\newcommand{\p}{$\pi$}
\newcommand{\mgbtwo}{MgB$_2$}
\newcommand{\tc}{$T_{\rm c}$}
\newcommand{\gkkp}{|g^{nn'}_{{\bf k,k'},\nu}|^2}

\newcommand{\up}{\uparrow}
\newcommand{\down}{\downarrow}

\newcommand{\I}{{\rm i}}
\newcommand{\E}{{\rm e}}

\newcommand{\Tr}[1]{{\rm Tr}\{ #1 \}}
\newcommand{\op}[1]{{\hat #1}}

\newcommand{\mint}[1]{\int\! {\rm d}^{3} #1 \, }
\newcommand{\mdint}[2]{\mint{#1}\!\!\!\mint{#2}}
\newcommand{\KS}{{\rm s}}

\newcommand{\Dk}{\Delta_{n{\bm k}}}
\newcommand{\Dkf}{\Delta_{n{\bm k_F}}}
\newcommand{\bkf}{{\bm k_F}}
\newcommand{\cabesi}{CaBeSi}
\newcommand{\cabesix}{CaBe$_x$Si$_{2-x}~$}
\newcommand{\albtwo}{AlB$_2$}
\newcommand{\pa}{$\pi_a$}
\newcommand{\pb}{$\pi_b$}

\begin{document}

\title{Multiband superconductivity in Pb, H under pressure and CaBeSi from {\it ab-initio} calculations}

\author{C. Bersier$^{3,4}$,  A. Floris$^{3,4}$, P. Cudazzo$^2$, G. Profeta$^2$,  A. Sanna$^{1,3,4}$, F. Bernardini$^1$, M. Monni$^1$, S. Pittalis$^{3,4}$, S. Sharma$^{3,4}$, H. Glawe$^{3,4}$, A. Continenza$^2$, S. Massidda$^1$, E.K.U. Gross$^{3,4}$}

\affiliation{$^1$SLACS-INFM/CNR,  and Dipartimento di Scienze Fisiche, Universit\`a degli Studi di Cagliari, I-09042
Monserrato (CA), Italy}
\affiliation{$^2$ CNISM - Dipartimento di Fisica,
Universit\`a degli Studi dell'Aquila, Via Vetoio 10,
I-67010 Coppito (L'Aquila) Italy}
\affiliation{$^3$Institut f{\"u}r Theoretische Physik, Freie Universit{\"a}t
Berlin, Arnimallee 14, D-14195 Berlin, Germany}
\affiliation{$^4$European Theoretical Spectroscopy Facility-ETSF}

\begin{abstract}
Superconductivity in Pb, H under extreme pressure and CaBeSi, in the framework of the density functional theory for superconductors, is discussed. A detailed analysis on how the electron-phonon and electron-electron interactions combine together to determine the superconducting gap and critical temperature of these systems is presented. Pb, H under pressure and CaBeSi are multigap superconductors. We will address the question under which conditions does a system exhibits this phenomenon. The presented results contribute to the understanding of multiband and anisotropic superconductivity, which has received a lot of attention since the discovery of MgB$_2$, and show how it is possible to describe   the superconducting properties of real materials on a fully {\it ab-initio} basis. 
\end{abstract}
\pacs{}
\maketitle


\section{Introduction}
The possibility of multiband superconductivity was proposed \cite{suhl}  
a few years after the formulation of BCS theory and was later experimentally 
observed\cite{shen,radebaugh,binnig}. In this regard, the details of the band ($n$) and ${\bf k}$ 
dependence of the superconducting (SC) gap were analyzed for many materials like 
Pb \cite{bennett,carbotte,wolfe} and Sn \cite{tin}. The recent discovery of
two-gap superconductivity in MgB$_2$ has renewed interest in this
phenomenon. Theoretical and experimental investigations have emphasized 
the importance of multiple gaps in the enhancement of the critical temperature \tc\ .
Theoretical estimates have shown that including the information about 
the $\sigma,\pi$ band dependence of the electron-phonon (e-ph) interaction in MgB$_2$  
greatly  enhances the calculated \tc\ and improves the agreement with experiment, 
as compared to an averaged, single-band, calculation\cite{liutwobands,choi,choinat,noimgb2}. 
Therefore of crucial importance is the question: 
When does the presence of multiple gaps occur and how does it affect \tc? In 
this context, it is very interesting to investigate the different materials showing this peculiarity. 

Recently, a novel approach to superconductivity, based on density 
functional theory (SCDFT)\cite{noiI,noiII}  has been able  to describe, in a completely 
parameter-free fashion, the superconducting properties
of several materials, ranging from the weak and intermediate to  strong coupling 
regime \cite{noiI,noiII,noimgb2} and from ambient to high pressure
conditions \cite{lial,K}. Unlike in the Eliashberg theory, in SCDFT 
the Coulomb interaction is treated on the same footing as the 
electron-phonon interaction. This is crucial for determining \tc\ accurately.

In this work, we solve the SCDFT gap equation in  a fully ($n,\bk)$-resolved formalism and 
calculate from first-principles the SC properties of Pb, H under pressure and CaBeSi . Within this approach  all 
the features of the SC gap emerge naturally, without any {\it a priori}, 
material specific physical model or adjustable parameters.  
Our results confirm the experimental finding that Pb is a two band superconductor.
The calculated  gaps, their overall anisotropy, and the \tc\ result in good 
agreement with experiments reported in Refs.~\onlinecite{blackford,lykken}.
A \tc\ enhancement of $\approx 8\%$ can be directly related to the presence of multiple gaps~\cite{Pb}. 

The SCDFT, with its truly  {\it ab-initio} character, allows not only 
to provide a better understanding of experimental facts, but also and more importantly, to predict the superconducting nature of systems which are
not yet experimentally realized. One prime example among those is H under extreme 
pressure. The possibility of high-temperature superconductivity in metallic 
hydrogen was suggested in 1968 by N. W. Ashcroft \cite{ashcroft}. Since then it has represented 
one of the most fascinating and intriguing topics involving fundamental issues
ranging from the limits of phonon mediation in the superconducting phenomenon, 
to implications in astrophysics\cite{Ash3}. In the following we show that 
SCDFT predicts electron-phonon mediated superconductivity in H, with \tc\ values up to 242 K at 450 GPa.

Another system for which SCDFT predicts multigap SC is CaBeSi~\cite{cabesi}, a material isostructural isoelectronic to  MgB$_2$.  Despite the many similarities with \mgbtwo,  according to our calculations \cabesi\ has  a very low  critical temperature (\tc$\approx 0.4$~K), consistent with experiment~\cite{sano}. However, \cabesi\  exhibits a complex gap structure, with three distinct gaps at the Fermi level.

The  paper is organized as follows: In Section \ref{scdft} we
summarize the main features of SCDFT and describe our computational approach; 
in Sections \ref{results_Pb}, \ref{results_H} and \ref{CaBeSi} we present our results for Pb, H under pressure and CaBeSi respectively; finally,  in Section \ref{concl}, we summarize our conclusions.

 \section{The Density functional theory for superconductors}
\label{scdft}

Density functional theory~\cite{DreizlerGross} has enjoyed increasing popularity 
as a reliable and relatively inexpensive tool to describe real materials.
In this section we will briefly outline the DFT approach to superconductivity, 
and refer to the original papers for more details. In order to give an introduction
to SCDFT, it is  instructive to recall how magnetism is treated within  DFT.
The Hohenberg-Kohn (HK) theorem~\cite{HohenbergKohn} states that
{\it all} observables, in particular also the magnetization, are functionals of 
the electronic density {\it alone}. This, however, assumes the knowledge of the 
magnetization as a functional of the density. Finding an approximation for 
this functional is extremely hard and, in practice, one chooses a different approach. 
The task can be vastly simplified by treating the magnetization density ${\bf m}(\br)$,
i.e., the order parameter of the magnetic state, as an
additional fundamental density in the density functional framework \cite{hedinvb}. 
An auxiliary field -- here a magnetic field ${\bf B}_{\rm ext}(\br)$ -- is introduced, 
which couples to ${\bf m}(\br)$ and breaks the corresponding (rotational) symmetry 
of the Hamiltonian. This field drives the system into the ordered state. If the 
system is actually magnetic, the order parameter will survive when the auxiliary
perturbation is quenched.
In this way, the ground-state magnetization density is determined by minimizing
the total energy functional (free energy functional for finite temperature
calculations) with respect to both the normal density and the magnetization
density.  Within this  approach much simpler approximations to the xc functional 
(now a functional of two densities) can lead to satisfactory results.

The same idea is also at the heart of density functional theory for
superconductors, as formulated by Oliveira, Gross and Kohn~\cite{ogk}.  Here the
order parameter is the so-called anomalous density,
\begin{equation}
  \label{eq:def-chi}
  \chi(\br,\br') = \Ave{\hat{\Psi}_{\up}(\br) \hat{\Psi}_{\down}(\br')} \, ,
\end{equation}
and the corresponding potential is the non-local pairing potential
$\Delta(\br,\br')$. It can be interpreted as an external pairing field, induced
by an adjacent superconductor via the proximity effect. Again, this external
field only acts to break the symmetry (here the gauge symmetry) of the system,
and is quenched at the end of the calculation. As in the case of magnetism,
if the system is actually a superconductor the order parameter will be sustained
by the self-consistent effective pairing  field.
The approach outlined so far captures, in principle, all the electronic degrees of
freedom.  To describe conventional phonon-mediated superconductors, also the
electron-phonon interaction has to be taken into account.

In  order to treat both weak and strong
electron-phonon coupling, the electronic and the nuclear degrees of freedom
have to be treated on equal footing. This can be achieved by a multi-component
DFT, based on both the electronic density and the diagonal of the nuclear N-body density matrix~\cite{kreibich,kreibich2008}~\footnote{Taking only the nuclear density would lead to a system of strictly
  non-interacting nuclei which obviously would give rise to non-dispersive,
  hence unrealistic, phonons.}:

\begin{equation}
  \label{eq:def-Gamma}
  \Gamma(\underline{\bR}) = 
  \Ave{\hat\Phi^\dagger(\bR_1) \dots \hat\Phi^\dagger(\bR_N)
    \hat\Phi(\bR_N) \dots \hat\Phi(\bR_1)},
\end{equation}
where $\hat{\Phi}(\bR)$ is a nuclear field operator. The quantity $\Gamma(\underline{\bR})$ is then included as an additional ``density'' in the formalism, besides the electronic density and the SC order parameter.

In order to formulate a Hohenberg-Kohn theorem for this system, we introduce a
set of three potentials, which couple to the three densities described above. Since
the electron-nuclear interaction, which in conventional DFT constitutes the external
potential, is treated explicitly in this formalism, it is {\em not} part of the
external potential. The nuclear Coulomb interaction $\hat{U}^{\rm nn}$ already has the 
form of an external many-body potential, coupling to $\Gamma(\underline{\bR})$, and
for the sake of the Hohenberg-Kohn theorem, this potential will be allowed to take
the form of an arbitrary N-body potential.
All three external potentials are merely mathematical
devices, required to formulate a Hohenberg-Kohn theorem. At the end of the
derivation, the external electronic and pairing potentials will be set to zero while 
the external nuclear many-body potential is set to  the nuclear Coulomb interaction.

As usual, the Hohenberg-Kohn theorem guarantees a one-to-one mapping between the
set of the densities $\{n(\br),\chi(\br,\br'),\Gamma(\underline{\bR})\}$ in
thermal equilibrium and the set of their conjugate potentials $\{v_{\rm
  ext}^\text{e}(\br)-\mu,\Delta_{\rm ext}(\br,\br'),v_{\rm
  ext}^\text{n}(\underline{\bR})\}$.  Therefore all the observables are
functionals of the set of densities. Finally, it assures that the grand canonical
potential,
  \be
  \label{eq:intomega}
  \Omega[n,\chi,\Gamma] = F[n,\chi,\Gamma] + \mint{r} n(\br) 
  [v^\text{e}_{\rm ext}(\br) - \mu]    
  - \mdint{r}{r'} \left[ \chi(\br,\br') \Delta_{\rm ext}^*(\br,\br') + \text{h.c.} \right]  
  + \mint{\underline R} \, \Gamma(\underline{\bR}) v_{\rm ext}^\text{n}(\underline{\bR})
  ,
  \ee
is minimized by the equilibrium densities. We use the notation $A[f]$ to denote that
$A$ is a functional of $f$. The functional $F[n,\chi,\Gamma]$ is
universal in the sense that it does not depend on the external potentials. It
is defined by
\be
  \label{eq:intF}
  F[n,\chi,\Gamma] = T^\text{e}[n,\chi,\Gamma] + T^\text{n}[n,\chi,\Gamma]   
  + U^\text{en}[n,\chi,\Gamma] + U^\text{ee}[n,\chi,\Gamma]
  - \frac{1}{\beta} S[n,\chi,\Gamma]
  \,,
\ee
where $T^\text{e}$ represents the electronic kinetic energy, $U^\text{ee}$ the electron-electron interaction, $T^\text{n}$ the nuclear kinetic energy, $U^\text{nn}$ the Coulomb repulsion between nuclei, $U^\text{en}$ the electron-nuclei interaction and $S$ is the entropy of the system,
\begin{equation}
  S[n,\chi,\Gamma] = -\Tr{\op\rho_0[n,\chi,\Gamma] \ln(\op\rho_0[n,\chi,\Gamma])}
  \,.
\end{equation}

In standard DFT one normally defines a Kohn-Sham system, i.e., a non-interacting
system chosen such that it has the same ground-state density as the interacting
one. The variational procedure for this system gives   Schr{\"o}dinger-like (Kohn-Sham) 
equations for non-interacting electrons subject to an effective (Kohn-Sham) potential.
These equations are nowadays routinely solved by solid state theorists. 
In our  formalism, the Kohn-Sham system consists of non-interacting
(superconducting) electrons, and {\it interacting} nuclei. We will not  describe
here the details of the method, and will only outline its basic features:
The Kohn-Sham potentials, which are derived in analogy to normal DFT, include
the external fields, Hartree, and exchange-correlation terms. The latter account
for all many-body effects of the electron-electron and electron-nuclear
interactions. Obtaining  their explicit form has represented a major theoretical 
effort\cite{KurthPhd,LuedersPhd,MarquesPhd}. 
Once this problem has been solved, 
the problem of minimizing the Kohn-Sham grand canonical
potential
 can be transformed into a set of three
differential equations that have to be solved self-consistently: One equation
for the nuclei, which resembles the familiar nuclear Born-Oppenheimer equation,
and two coupled equations which describe the electronic degrees of freedom and
have the algebraic structure of the Bogoliubov-de Gennes~\cite{Bogoliubov58}
equations.

The resulting Kohn-Sham Bogoliubov-de Gennes (KS-BdG) equations read (we use atomic Hartree units)

\begin{subequations}
  \label{KS-BdG}
  \begin{align}
    \left[ -\frac{\nabla^2}{2} + v^\text{e}_\KS(\br) - \mu \right]
    u_\nk(\br)   
    +\mint{r'} \Delta_\KS(\br,\br') v_\nk(\br') &= \tilde{E}_\nk \, u_\nk(\br) \,,   \\
    - \left[ -\frac{\nabla^2}{2} + v^\text{e}_\KS(\br) - \mu \right]
    v_\nk(\br) 
    + \mint{r'} \Delta^*_\KS(\br,\br') u_\nk(\br') &= \tilde{E}_\nk \, v_\nk(\br)\,, 
  \end{align}
\end{subequations}
where $u_\nk(\br)$ and $v_\nk(\br)$ are the particle and hole amplitudes.  This
equation is very similar to the Kohn-Sham equations in the OGK
formalism~\cite{ogk}.  However, in the present formulation the lattice potential
is not considered an external potential but enters via the electron-ion
Hartree term. Furthermore, our exchange-correlation potentials depend on the
nuclear density matrix, and therefore on the phonons. Although Eq.
 (\ref{KS-BdG}) and 
 the corresponding equation for the nuclei
have the structure of static mean-field equations, they contain, in principle,
all correlation and retardation effects through the exchange-correlation
potentials.  

These KS-BdG equations can be simplified by the so-called decoupling
approximation~\cite{GrossKurth91,noiI}, which corresponds to 
approximating the particle and hole amplitudes by:
\begin{equation}
  \label{eq:decapprox}
  u_\nk(\br) \approx u_\nk \varphi_\nk(\br)
  \,; \quad
  v_\nk(\br) \approx v_\nk \varphi_\nk(\br)
  \,,
\end{equation}
where the wave functions $\varphi_\nk(\br)$ are the solutions of the normal
Schr{\"o}dinger equation.  In this way the eigenvalues in Eqs.~(\ref{KS-BdG})
become $\tilde E_\nk = \pm E_\nk$, where
\begin{equation}
  E_\nk = \sqrt{\xi_\nk^2+|\Delta_\nk|^2} \,,
\end{equation}
and $\xi_\nk = \epsilon_\nk-\mu$. This form of the eigenenergies allows us to
interpret the pair potential $\Delta_{\nk}$ as the gap
function of the superconductor. Furthermore, the coefficients $u_\nk$ and $v_\nk$ 
are given by simple expressions within this approximation
\begin{subequations}
  \begin{align}
    u_\nk & = \frac{1}{\sqrt{2}}{\rm sgn}(\tilde E_\nk) \E^{\I\phi_\nk}
    \sqrt{1+\frac{\xi_\nk}{\tilde E_\nk}} \,,
    \\
    v_\nk & = \frac{1}{\sqrt{2}} \sqrt{1-\frac{\xi_\nk}{\tilde E_\nk}}
    \,.
  \end{align}
\end{subequations}
Finally, the matrix elements $\Delta_\nk$ are defined as
\begin{equation}
  \label{eq:delta_i}
  \Delta_\nk = \mdint{r}{r'} \varphi^*_\nk(\br)\Delta_\KS(\br,\br')\varphi_\nk(\br')
  \,,
\end{equation}
and $\phi_\nk$ is the phase $\E^{\I\phi_\nk} = \Delta_\nk/|\Delta_\nk|$.  The
normal and the anomalous densities can then be easily obtained from:
\begin{subequations}
  \label{eq:el-densities}
  \begin{align}
    n(\br) & = \sum_\nk \left[1-\frac{\xi_\nk}{E_\nk}
      \tanh\left(\frac{\beta}{2}E_\nk \right)\right]
    |\varphi_\nk(\br)|^2
    \\
    \chi(\br,\br') & = \frac{1}{2}\sum_\nk
    \frac{\Delta_\nk}{E_\nk} \tanh\left(\frac{\beta}{2}E_\nk \right)
    \varphi_\nk(\br)\varphi^*_\nk(\br')
    \,.
  \end{align}
\end{subequations}
Within the decoupling approximation, we finally arrive at an equation for the 
 {\bf k}-resolved superconducting gap $\Delta_{n{\bf k}}$, which  has the following form\cite{ogk,noiI,noiII,PsikRev}:
\begin{equation}
  \label{eq:gap} 
  \Delta_\nk = - {\cal Z}_\nk \Delta_\nk -\frac{1}{2}\sum_\nkp
  {\cal K}_{\nk,\nkp} \frac{\tanh\left(\frac{\beta}{2}E_\nkp\right)}{E_\nkp} 
  \Delta_\nkp
  \,.
\end{equation}

Eq. (\ref{eq:gap}) is the central equation of the DFT for superconductors. The kernel
 ${\cal K}$  consists of two contributions 
${\cal K}={\cal K}^{\rm e-ph}+{\cal K}^{\rm e-e}$, representing the effects of 
the e-ph and of the e-e interactions, respectively. The diagonal term ${\cal Z}$ plays a similar role as the renormalization term in the Eliashberg 
equations. Explicit expressions  
of ${\cal K}^{\rm e-ph}$ and ${\cal Z}$, which are the results of
the approximate functionals, are given in Eqs. 9 and 11 of Ref.~\cite{noiII} 
respectively. These two terms
involve the e-ph coupling matrix, while ${\cal K}^{\rm e-e}$ contains the matrix elements 
of the screened Coulomb interaction. Eq.~(\ref{eq:gap}) has the same structure 
as the BCS gap equation, with the kernel ${\cal K}$ replacing the model interaction 
of BCS theory. This similarity allows us to interpret the kernel as an effective 
interaction responsible for the binding of the Cooper pairs.  Moreover, we emphasize that Eq.~(\ref{eq:gap}) is not a mean-field equation (as  
in BCS theory),  since it contains correlation effects via the SC exchange-correlation functional entering ${\cal K}$ and ${\cal Z}$. Furthermore, it has the 
form of a static equation -- i.e., it does not depend {\it explicitly} on the frequency --  
and therefore has a simpler  (and computationally more manageable) structure  than the Eliashberg equations. However, this 
certainly does not imply that retardation effects are absent from the theory.  Once again, retardation effects
enter through the xc functional, as explained in Refs. ~\cite{noiI,noiII} . The SCDFT allows to treat the e-ph and the screened e-e interactions  on the same footing.
These terms, however, can be treated at different  levels of approximation.  

 We calculated the screened Coulomb matrix elements (ME) with respect to the Bloch functions, for the whole energy range 
of relevant valence and conduction states.  The different 
nature of the  electronic  bands in each material ($e.g.$  some of them can be highly localized 
while others more delocalized), strongly calls for the use of a non-diagonal screening, 
including local field effects.  

As mentioned above, the  normal state calculations,  necessary for the study  of the 
superconducting state,  are performed within DFT in the LDA or GGA approximations.
Computationally, the electronic and  phononic properties are obtained using the
pseudopotential method as implemented in the QUANTUM-ESPRESSO package\cite{pwscf};
 the screened Coulomb matrix elements are obtained with the SELF\cite{SELF} code.


\section{Multiband superconductivity in $\rm Pb$}
\label{results_Pb}

\begin{figure}[t]
\begin{center}
\includegraphics[clip,width=0.43\textwidth]{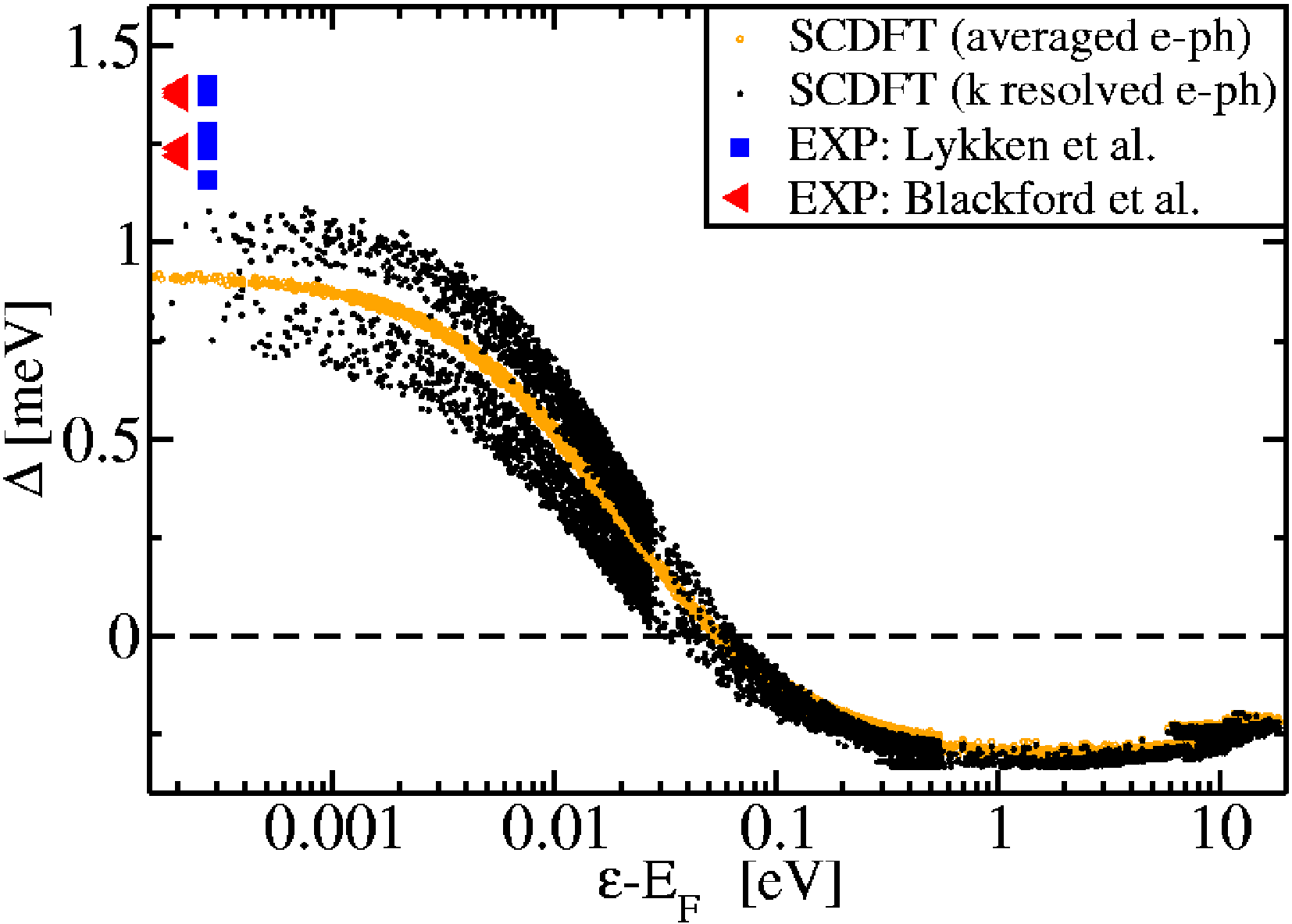}
\end{center}
\caption{Superconducting gap of Pb as a function of the energy distance from
the Fermi energy ($T=0$~K). Small black points: \D~ resulting from the inclusion 
of the ${\bf k}$-resolved e-ph and Coulomb matrix elements. Small orange points: \D~ calculated with a corresponding averaged  e-ph interaction
(Eliashberg function $\aF$). Big triangles and squares: experimental 
values (from Refs.~~\onlinecite{blackford}, \onlinecite{lykken}).}
\label{fig1}
\end{figure}

The peculiar aspects of superconductivity in Pb have been
discussed in the past
\cite{horowitz,vanderhoeven,overhouser,bennett,blackford,lykken,carbotte,wolfe}. 
In this section we present the results for the SC properties of Pb, calculated within the 
SCDFT formalism and highlight the importance of the ($n,\bk)$ gap anisotropy in this system. We first concentrate on the SC gap 
function, i.e. the solution of Eq. (\ref{eq:gap}). There are two ways to visualize 
this function: The first is to look at the gap as a function of the normal-state energy eigenvalue. 
The set of black points in 
Fig. \ref{fig1} shows the gap at $T=0$~K calculated on a set of random {\bf k} points, as a 
function of the energy distance from the Fermi level \EF.  An important feature
of this plot is that for each energy the gap is not a single-valued function. 
This means that, in general, the SC gap is not isotropic in reciprocal space, 
i.e.  for $\varepsilon=$\EF~ its value depends on the Fermi vector ${\bf k}_F$. 
In particular, for each $\varepsilon\approx$ \EF~ we observe two distinct ``sets'' 
of gaps, in strict analogy with the case of MgB$_2$  (see, for comparison, 
Fig.1 in Ref.~~\onlinecite{noimgb2}). Moreover, each ``set'' has an associated, finite, vertical energy spread. 

Most importantly the gap function has a negative 
tail extending to high energy. This is a general feature of the gap function, due to the presence of the Coulomb repulsion and to its dominance for large energies. Due to this feature, Coulomb renormalization effects are properly included in the SCDFT formalism. We emphasize that these effects are crucial to obtain a predictive superconducting solution.

\begin{figure}
\includegraphics[width=0.7\textwidth]{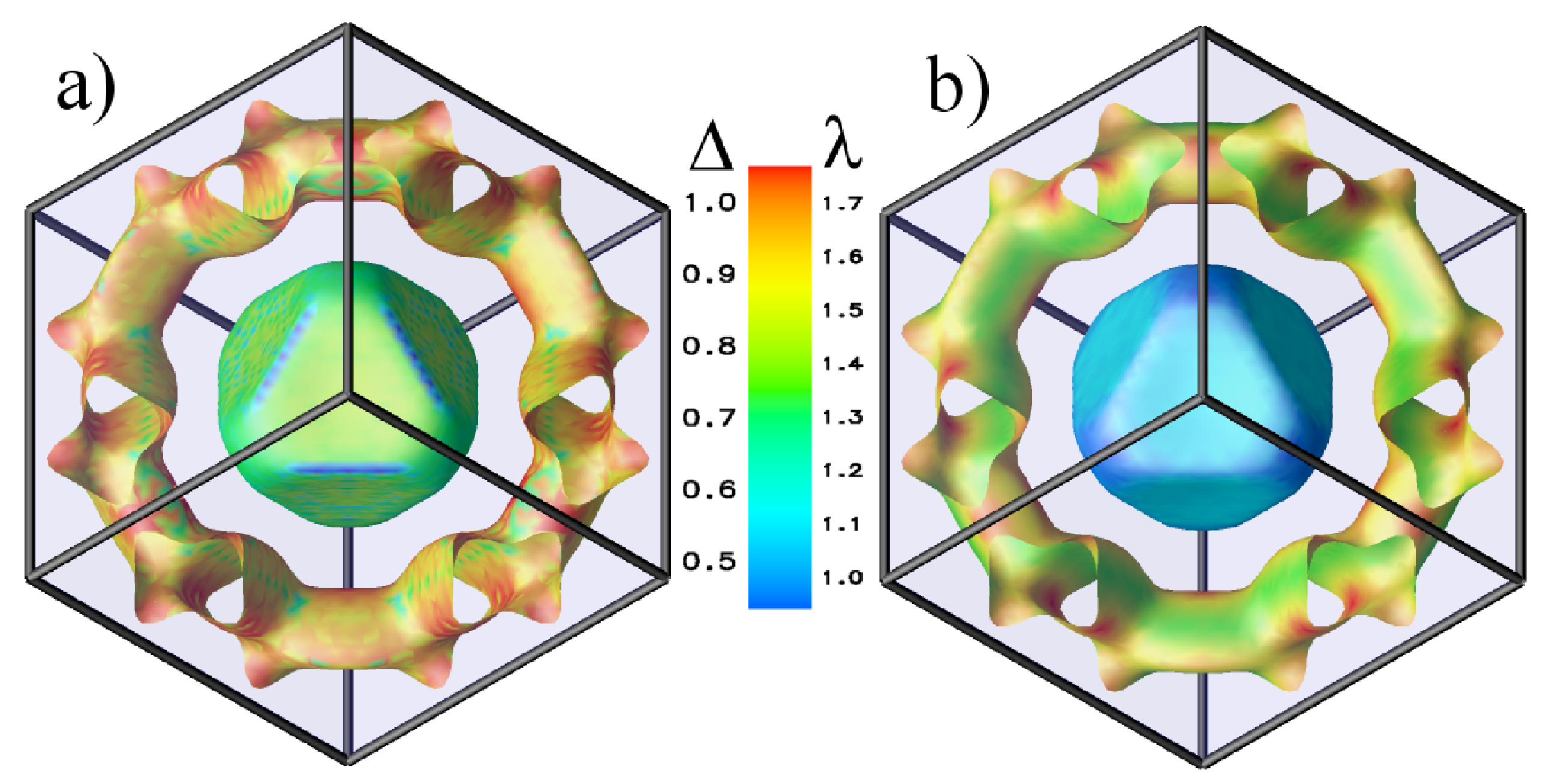}
\caption{a) (left): Superconducting gap $\Dk$~ calculated at the Fermi surface of Pb, at $T=0$\, K. b) (right): Electron-phonon interaction $\lambda_{n\bk}$ at FS.} 
\label{fig2}
\end{figure}

A second possibility to visualize the gap is to define equi-energy surfaces and 
plot the gap value on that surface. The equi-energy surface can be chosen to be
the Fermi surface (FS). 
In  Fig. \ref{fig2}a we plot the FS of Pb, consisting of two separate 
sheets, coming from the two (essentially  $p$) bands crossing \EF. The sheets
are topologically quite different, the first one being rather spherical and the
second having a more complex tubular-like structure. 
The color\s in  Fig. \ref{fig2}a represent the values  of $\Delta_{n{\bm k}_F}$. 
It is clear that the two distinct sets of \D\ in Fig. \ref{fig1} come from the 
two sheets of the Fermi surface.  
Fig. \ref{fig1} shows that their energy separation is in good agreement
with the experiments \cite{lykken,blackford}.

Fig. \ref{fig2}a also shows that $\Dk$ is anisotropic inside each single sheet, 
resulting in the vertical energy spread mentioned before (intraband gap anisotropy).
Continuing our analysis, in Fig. \ref{fig2}b we plot  the e-ph coupling 
$\lambda_{n{\bm k_F}}$  on the FS, where
\be
\lambda_{n{\bm k}}=2\sum_{\nkp,\nu}\frac{\gkkp}{\omega_{{\bm k'-\bm k},\nu}} \times  \delta(\varepsilon_{\nkp}-E_F)
\ee 
is the average of all the possible e-ph scattering processes connecting 
two points at FS, but always involving the electronic initial state $(n,{\bm k})$. 
The total e-ph coupling constant  $\lambda$ can be expressed as 
$\lambda=\frac{1}{N(E_F)}\sum_{\nk}\lambda_{n\bk}\times \delta(\varepsilon_{\nk}-E_F) $. 
Comparing the plots in Fig. \ref{fig2}a and Fig. \ref{fig2}b we notice a 
striking similarity, that indicates a strong correlation between the $n$-dependence 
and ${\bf k}$ anisotropy of the e-ph coupling  and of the gap. This is a
general feature that we also observe for other materials. Moreover, our calculations
show that the gap anisotropy is due  to the e-ph coupling only: the open orange symbols  in Fig. \ref{fig1} shows the gap function 
calculated with an average e-ph interaction, i.e. including the phononic
kernels ${\cal K}^{\rm e-ph}$ and ${\cal Z}$ that contain the Eliashberg function 
$\aF$ (see Eqs. 23 and 24 of Ref.~ \onlinecite{noiII}), but still with the 
Coulomb ME~ ${\cal K}_{\nk,\nkp}^{\rm e-e}$. It is clear that the e-ph 
average washes out all the band and ${\bm k}$ dependence of the
gap (i.e. $\Dk=\Delta(\varepsilon_{n{\bm k}})$). 

Eq. (\ref{eq:gap}) allows to calculate  $\Dk$ as a function of temperature $T$, 
defining \tc\ from the condition that $\Dk(T_c)=0$. We obtain \tc  $=5.25$~K 
and \tc $=4.84$~K for the anisotropic and average calculation respectively. 
We see that, although certainly not in a proportion comparable with MgB$_2$, 
the presence of an anisotropic, multiband gap produces an $8\%$ enhancement of 
the value of \tc~. In MgB$_2$ this effect is much more pronounced
\cite{liutwobands,choi,choinat,noimgb2} than in the present case. This is partially 
due to the much lower difference in e-ph coupling between the two bands 
(in MgB$_2$ the coupling in the $\sigma$ bands is roughly 3 times higher than 
in the $\pi$ bands).

An interesting question is under which conditions does a system show multiband
superconductivity, i.e. with a clear separation between the gaps. Looking again
at the $\lambda_{n\bkf}$ we see that in Pb the two sets corresponding to the two 
bands crossing \EF~ are not continuous, i.e. the values of $\lambda_{1\bkf}$  
and $\lambda_{2\bkf}$ do not superimpose. In order to check whether the presence 
of two separate gaps is related directly to this property of the $\lambda_{n\bkf}$,
we performed several model calculations, rescaling  the e-ph ME, in order to 
have a continuous set of  $\lambda_{n\bkf}$ {\it and} keeping constant the value 
of the total, average $\lambda$. Most interestingly we found that the two SC 
gaps $\Dkf$ get continuous exactly when the two set of $\lambda_{n\bkf}$ do, 
independent of the details of how the coupling were modified to bring 
the $\lambda_{n\bkf}$ to merge. This leads us to conclude that Pb is a two gap 
system due to the fact that
$\lambda_{n\bkf}$ are disjoint sets (relative to the band $n$). In general 
there are two necessary conditions to fulfill in order to obtain multigap 
superconductivity. First the system needs to have disconnected FS sheets and
second the electron-phonon coupling must be different on each of these sheets.


\section{Superconductivity in $\rm H$ under pressure}\label{results_H}


\begin{figure} 
\includegraphics[width=0.4\linewidth]{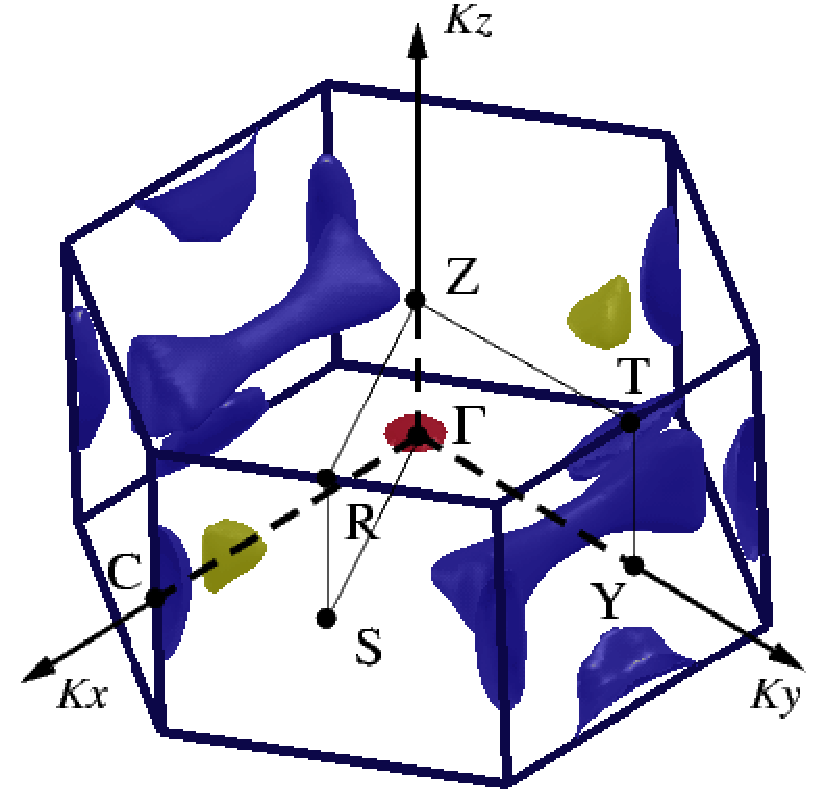}
\caption{Fermi surface at 414 GPa. 
Different colors represents different values
of the superconducting gap on the FS. 
Red ($18.0<\Delta<21 \rm\ meV$), 
yellow ($13.6<\Delta<18.0 \rm\ meV$) and blue ($10.0<\Delta<13.6 \rm\ meV$).}
\label{fig3}
\end{figure}

Based on the simple BCS
theory of superconductivity, we can understand why molecular metallic hydrogen could be
a good superconductor\cite{Ash2}: This system has very high phonon 
frequencies due to the light H mass, and it has a possibly strong electron-phonon
interaction related to the lack of core electrons and to the quite strong covalent 
bonding within the H$_2$ molecules. Many studies, aiming at investigating further this 
possibility \cite{Ash4,Cohen,Zhang}, collected strong evidence pointing to
high-Tc superconductivity. At present, however, the full scenario is still far 
from being clear and well established.

The low temperature and 
high pressure ($>$ 400 GPa) phase of molecular hydrogen is predicted to be a
base-centered orthorhombic metallic molecular solid (known as $Cmca$ phase \cite{needs}) 
with two molecules per unit cell located on different layers. The electronic 
band structure arises from the bonding and anti-bonding
combination of the H$_2$ molecular orbitals. At high pressure, the band overlap 
between the valence and conduction bands produces a rich and complex Fermi surface (Fig. \ref{fig3}) with  disconnected sheets of different orbital nature\cite{noiH} and provides the 
strong e-ph coupling necessary to superconductivity (see below).

\begin{figure}[t]
\includegraphics[width=0.4\linewidth,clip=]{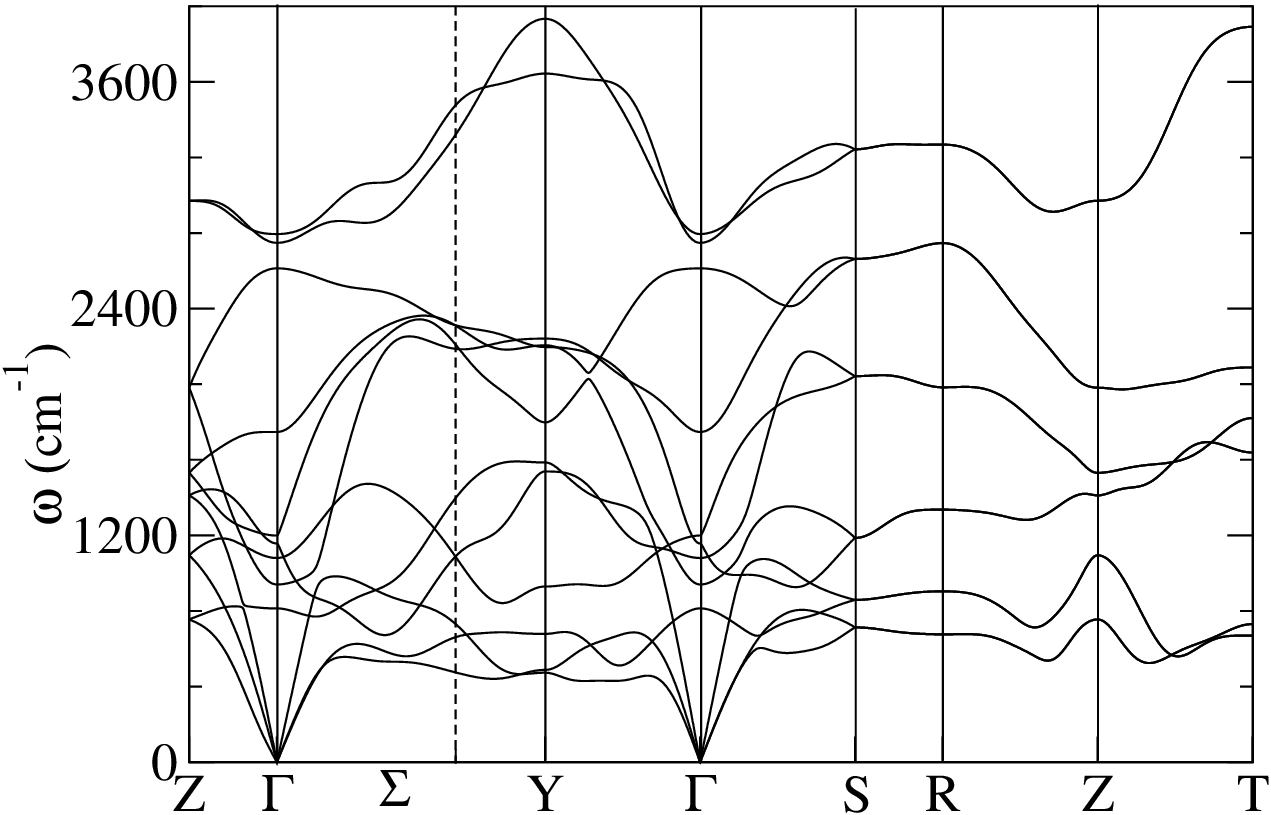}
\caption{H phonon dispersion at 414 GPa.}
\label{fig4}
\end{figure}
Three types of bands form the Fermi surface of H at high pressure: The two tubular
structures intersecting the $k_y$ axis and the two prism-like structures along the 
$k_x$ axis have hole-type character; the disk centered at the $\Gamma$-point and the structures near the C-point have electron-type character; the remaining sheets 
are related to degenerate orbitals along Z - T. The different orbital 
character of the FS branches suggests the occurrence of different couplings 
between the various bands, leading to an anisotropic SC gap and to  
multiband superconductivity. 

\begin{figure}[b]
\includegraphics[width=0.4\linewidth,clip=]{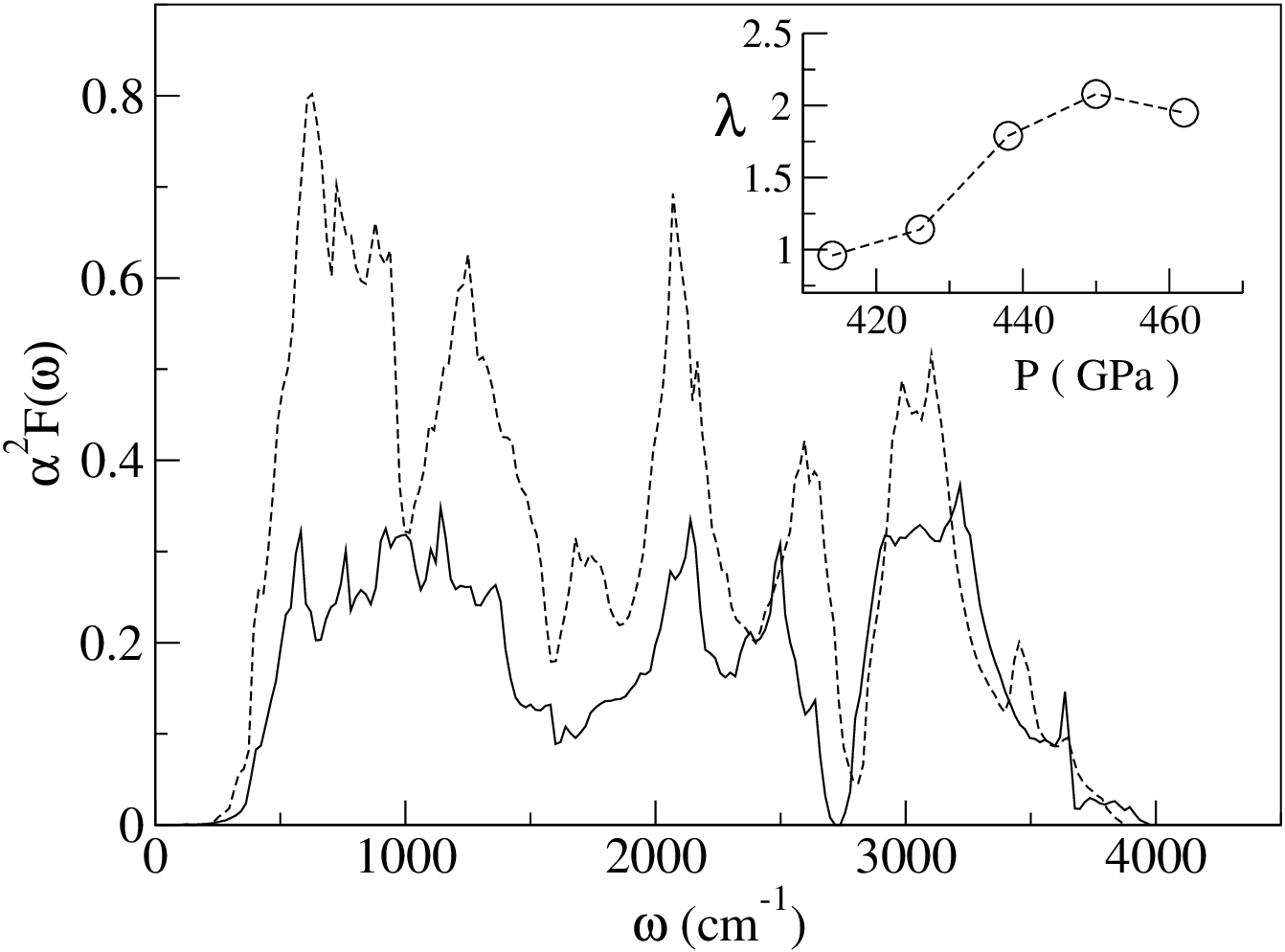}
\caption{$\alpha^2F(\Omega)$ at 414 (solid line) and 462 (dashed line) GPa. The
inset shows  the electron-phonon coupling ($\lambda$) 
as a function of the pressure.}
\label{fig5}
\end{figure}

The molecular nature of the charge density distribution gives rise to an intriguing phonon dispersion (Fig.\ref{fig4}). The phonon modes can be grouped into three main branches: phononic, libronic and vibronic, corresponding respectively to the relative translations, rotations, and internal vibrations of the H$_2$ molecules. The three FS regions possess distinct e-ph coupling: The most coupled one is the disk at $\Gamma$ with $1.8 < \lambda_{\bf k} < 2.00$; then, we find the two prism-like structures with $1.00 < \lambda_{\bf k} < 1.80$, and the other sheets with $\lambda_{\bf k} < 1.00$.  The tubular structures have the smallest $\lambda_{\bf k}$. The presence of multiple Fermi surfaces provides a "{\bf q}-distributed" coupling, with many modes contributing to the pairing. This allows to increase $\lambda$ still avoiding a lattice instability, which could result from very high  coupling at some specific $\bf{q}$'s. In fact, the Eliashberg function (Fig.\ref{fig5}) shows  three main frequency regions $all$ of them strongly coupled to the electrons and associated with phononic, libronic and vibronic modes, in an  increasing frequency order. However, a net distinction between phononic and libronic modes (in the low frequency region: 0 $\leq\omega\leq$ 1600 cm$^{-1}$) is prevented due to the branch mixing  caused by the large renormalization of the libronic modes. As the pressure raises from 414 GPa, the coupling increases and a new feature appears at $\approx$ 1700 cm$^{-1}$ (Fig. \ref{fig5}, plot at 462 GPa). The additional peak in the $\alpha^2F(\omega)$ comes from  an extra band which  crosses the  Fermi level close to the R-point. This creates a new electron-like Fermi sheet, strongly coupled with in-phase libronic modes, and opens a further coupling channel at higher pressure giving rise to a jump of $\lambda$ at $\approx$ 435 GPa (inset of Fig. \ref{fig5}).

Although rather large, these values of $\lambda$ are of the same order as, e.g., those obtained in fcc-Li under high pressure (40 GPa). There, $T_c$  reached $\simeq$ 20 K, a very large value for a free electron-like metal. In the case of hydrogen, we expect that the large phonon frequencies provided by the H$_2$ vibronic modes lead to much  higher $T_c$'s.

\begin{figure} 
\includegraphics[width=0.4\linewidth,angle=-90]{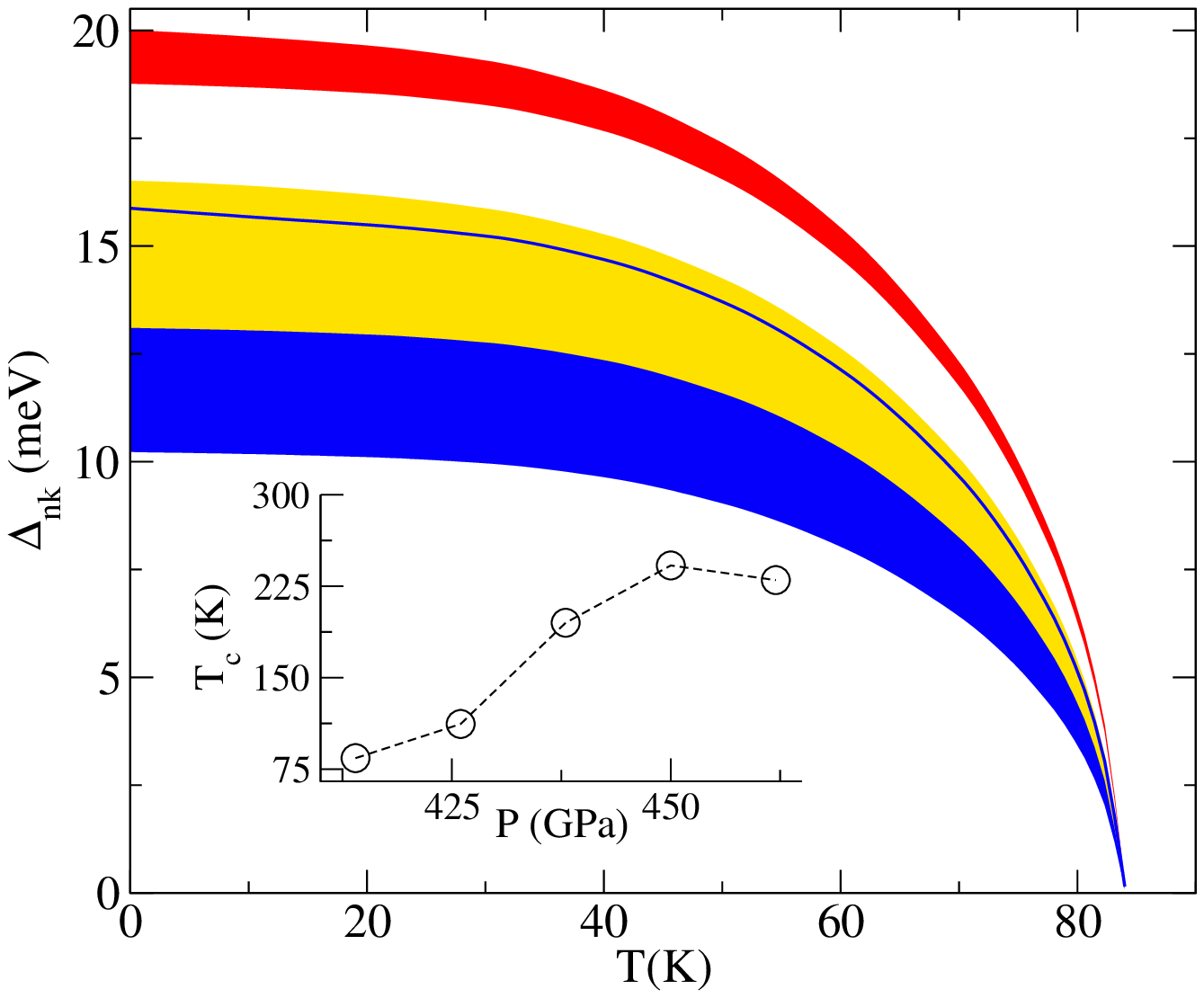}
\caption{$\Delta_{n{\bf k}}$ at E$_F$ at 414 GPa. 
The shaded regions represent the ${\bf k}$-anisotropy
over different bands (note that the blue and yellow gaps overlap from
13.00 to 15.75 meV). 
The inset shows the superconducting critical temperature 
as a function of  pressure.}
\label{fig6}
\end{figure}

The $T_c$  was calculated solving the SCDFT anisotropic gap equation Eq. (\ref{eq:gap}). As previously shown in the case of low-density metals\cite{lial} and discussed by Richardson and Ashcroft\cite{Ash5} a complete inclusion of both electron-phonon and electron-electron interactions is crucial to reproduce the experimental $T_c$'s. We find very high $T_c$'s, considerably increasing with pressure up to 242 K at 450 GPa (inset of Fig.\ref{fig6}). Comparing Figs. \ref{fig5} and \ref{fig6} insets, it is clear that $T_c$ follows
 closely the $\lambda$ behavior with pressure.

We predict a multigap (three gap) superconductivity, which is progressively lost at high pressure, where interband scattering raises the nondiagonal matrix elements, leading to a merging of the different gaps. The occurrence of multigap superconductivity is clearly illustrated in Fig. \ref{fig6}, which shows the $\Dk$ at the Fermi level as a function of temperature [for all the bands (n) and {\bf k}-points] and at the pressure with largest band anisotropy, P = 414 GPa (Tc = 84 K). The three gaps are evident: The largest one at $\approx$ 19.3 meV, and two very anisotropic, overlapping gaps at 15.4 and 13.6 meV. The three gaps are associated with the three different FS sheets (cf. Fig. \ref{fig3}): The strongly coupled disk around the $\Gamma$-point, the prism-like sheets, and others associated with the lowest gap. 

 In H the main source of band anisotropy is given by the e-ph interaction. Furthermore a simple multigap BCS model (not including anisotropy in $\mathbf{k}$) although not adequate to treat strong coupling system, is able to reproduce qualitatively the full SCDFT results~\cite{prbH}. This suggests that the dominant effect  of the anisotropy on \tc\ is related to the band contribution and that the {\bf k} space anisotropy is less relevant. We emphasize, however, that although the anisotropy of the interaction generates three different gaps, it does not play a fundamental role in enhancing $T_c$, which shows an increase of only 10K compared to the value obtained within an isotropic approximation. This is due to the fact that in H the inter-band (low-{\bf q}) coupling dominates over the intra-band coupling (high-{\bf q})~\cite{prbH}. This is different from \mgbtwo, where the intra-band coupling is much stronger and the anisotropy doubles the critical temperature~\cite{choi,noimgb2}.

\section{Superconductivity in $\rm CaBeSi$}
\label{CaBeSi}

\begin{figure}[t]
  \begin{center}
    \includegraphics[clip,width=0.4\textwidth]{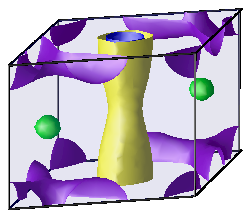}

  \end{center}

  \caption{\cabesi\ Fermi surface. The colors identify the three gaps at \EF: the largest \s\ (yellow and blue cylindrical-like sheets); the intermediate \p\ bonding (green spheres); the lowest \p\ antibonding  (violet).}
  \label{fig7}
\end{figure}

In this section we report the SC properties of \cabesix at $x=1$ (\cabesi\ in the following) in the {\it \albtwo} phase, with Be and Si atoms alternating within the honeycomb layers with an $AA$ stacking. A general similarity is found between  \cabesi\ and  \mgbtwo\ : in both materials the \s, $\pi$-bonding (\pb) and $\pi$-antibonding (\pa) bands cross the Fermi level (\EF).  In \cabesi\ the \pb\ bands are almost fully occupied, leaving only small hole pockets at K which give rise to the little \p\  spheres of the Fermi surface (Fig.~\ref{fig7}). These replace the \p\ tubular structure present in \mgbtwo.

 Despite the general similarities, \cabesi\ and \mgbtwo\
have rather different chemical and phononic properties, with important consequences for the e-ph interaction~\cite{cabesi}. 
In fact, \cabesi\  phonon frequencies are lower in comparison with \mgbtwo, mainly due to the larger mass of Ca and Si versus Mg and B. In the CaBeSi phonon dispersion, the E$_{2g}$ mode is fairly flat along the in-plane BZ symmetry 
 lines and it shows only a very weak renormalization (four times smaller than in \mgbtwo\ \cite{kong,bohnen,shukla}) due to the very small E$_{2g}$ electron-phonon matrix elements. In turn, this is related to the delocalized and ionic nature of the \s\ bonds in \cabesi\ . In fact, the connection between strongly covalent bonds  and strong e-ph coupling seems to be a general feature \cite{an-pickett,lial,K,Pb}. The small e-ph coupling makes \cabesi\ a weak coupling superconductor with e-ph $\lambda= 0.38$ and \tc$=0.4$~K, in spite of a nesting at the Fermi surface twice as big as in \mgbtwo\ .

\begin{figure}[t]
  \begin{center}
    \includegraphics[clip,width=0.43\textwidth]{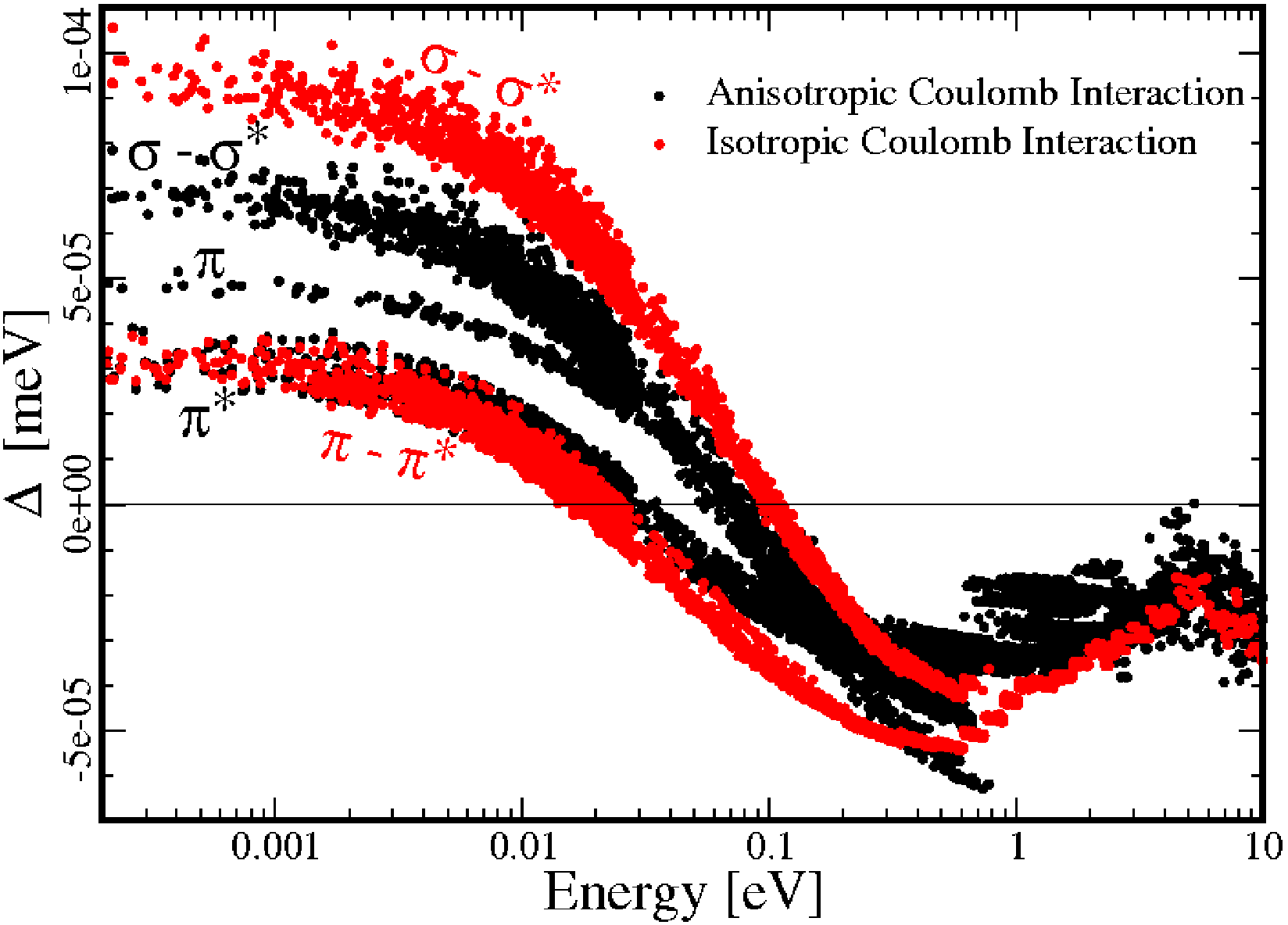}

  \end{center}

  \caption{\cabesi\ superconducting gap as a function of the energy distance from the Fermi level.}
  \label{fig8}
\end{figure}

 The solution of the self-consistent gap equation reveals an unexpected complex structure (Fig.~\ref{fig8}) with clearly separated three gaps at \EF\ (unlike in H where two of the three gaps superpose). The calculated critical temperature is very low (T$_c$= 0.4), lower than the upper limit (4.2 K) set by the experimental results\cite{sano}. Unlike in \mgbtwo, where  superconductivity is interpreted  within a two-band model\cite{liutwobands,brinkman,golubov,gonnelli_2gap}, in \cabesi\ there is a further  \pb-\pa\ gap splitting. As in \mgbtwo, the largest gap is related to the \s\ FS sheets (cylindrical-like structures in Fig.~\ref{fig7}), the intermediate one to \pb\ sheets (small hole spheres) and the lowest to \pa\ sheets. The additional \pb-\pa\ gap splitting is a peculiar feature of \cabesi\ not present in \mgbtwo, where the two \s\ and the two \p\ gaps merge together.
In order to understand the origin of this splitting we perform some additional computational experiments, solving the gap equation $(i)$ completely  neglecting the Coulomb interaction, $(ii)$ including only an averaged Coulomb term and $(iii)$ with isotropically averaged Coulomb and phononic interactions, corresponding to the dirty limit. In both $(i)$ and $(ii)$ cases, the three gap structure is destroyed, bringing back to a two-band, \mgbtwo-like gap structure. In case $(iii)$, instead, superconductivity is completely lost. As a result, we predict superconductivity in \cabesi\ {\it only} if the anisotropic structure of the interactions is included. While, as in \mgbtwo, the \s-\p\ gap splitting is related to the different e-ph coupling in these bands, the further \pb-\pa\ splitting  is a pure effect of the complex structure of the anisotropic Coulomb repulsion, acting  in a different way on the  \pb\ and \pa\ states.

\section{Summary}\label{concl}


Three multi-gap superconductors were investigated in this work:
Pb, H under pressure, and CaBeSi.
Without assuming any ad-hoc model or adjustable parameters, we find that
Pb is a two-gap superconductor, in agreement with available experiments
and
with previous theoretical results based on other methods.  We show that
the
$n$ and $\bk$ gap anisotropy and the separation of the two gaps correlate
strongly with the anisotropy of the
electron-phonon interaction on the different sheets of the Fermi surface.
This multi-gap character produces an enhancement of \tc\ relative to
the isotropic case.
 
 For H, we find a strong increase of the critical temperature with
 increasing pressure.
 As the corresponding pressure regime is not yet experimentally accesssible
 our
 calculation of the astonishingly high critical temperatures represents a
 genuine
 prediction. The analysis of the electronic and phononic properties points
 to the
 origins of the strong enhancement of superconductivity with rising
 presssure: In particular:
 $i)$ we confirm a strong electron-phonon coupling, increasing with
 pressure;
 $ii)$ the presence of multiple Fermi surfaces with different character
 provides both strong intra-band (low-{\bf q}) and inter-band (high-{\bf
 q})
 electron-phonon scattering. In this respect, $iii)$ a major role is played
 by the molecular rotational (libronic) and vibrational (vibronic) modes,
 which are strongly coupled with the inter-molecular charge. Finally,
 $iv$) we predict three superconducting gaps at the Fermi level, associated
 with Fermi surface branches having different character.
  
  We have also investigated the superconducting properties of \cabesi,
  a material isostructural and isoelectronic to \mgbtwo. While the band
  structures present strong similarities, the phonon structure and the e-ph
  interaction of the two systems differ substantially. In particular,  the
  less localized \s\ charge of \cabesi\ leads to a dramatic reduction in the
  $E_{2g}$ electron-phonon coupling, with a consequent reduction of the
  phonon
  renormalization seen in \mgbtwo. Interestingly, \cabesi\ exhibits three
  superconducting gaps at the Fermi level. The \pb-\pa\ splitting  is a pure
  effect of the complex structure of the anisotropic Coulomb repulsion,
  acting
  differently on the  \pb\ and \pa\ states.

    The role of multigap superconductivity in enhancing the critical
    temperature of
    superconducting materials has been known for a long time. However, the
    crucial challenge
    is to predict under which circumstances and for which materials the
    phenomenon of multi-gap
    superconductivity actually occurs. With the SCDFT approach which features
    a
    fully anisotropic description of the electron-phonon coupling and of the
    screened
    Coulomb repulsion on an {\it ab-initio} basis, this goal has been
    achieved.

\section{Acknowledgments}
This work makes use of results produced by the Cybersar Project 
managed by the Consorzio COSMOLAB,   
co-funded by the Italian Ministry of University and Research 
(MUR) within the Programma Operativo Nazionale 2000-2006 
"Ricerca Scientifica, Sviluppo Tecnologico, Alta Formazione" 
per le Regioni Italiane 
dellï¿½Obiettivo 1  ï¿½ Asse II, Misura II.2 ï¿½Societï¿½ dellï¿½Informazioneï¿½.
Work partially supported by the Italian Ministry of
Education, through PRIN 200602174 project, by INFM-CNR
through a supercomputing grant at Cineca (Bologna, Italy),
by the Deutsche Forschungsgemeinschaft within SPP1145 and by
NANOQUANTA Network of Excellence of the European Union.


\end{document}